\documentclass[12pt]{article}

\usepackage{amsmath,amsfonts,enumerate,array}
\usepackage{graphicx}
\usepackage[usenames]{color}
\usepackage[english]{babel}
\usepackage{fancybox}
\usepackage{chngpage} 

\newcommand{\captionfonts}{\small}

\makeatletter  
\long\def\@makecaption#1#2{%
  \vskip\abovecaptionskip
  \sbox\@tempboxa{{\captionfonts #1: #2}}%
 \ifdim \wd\@tempboxa >\hsize
    {\captionfonts #1: #2\par}
  \else
    \hbox to\hsize{\hfil\box\@tempboxa\hfil}%
  \fi
  \vskip\belowcaptionskip}
\makeatother   

\usepackage{mciteplus}

\usepackage[colorlinks=true,      linkcolor=blue,      urlcolor=blue,      filecolor=blue,      citecolor=blue,
      pdfstartview=FitH,     pdfpagemode=UseNone,      bookmarksopen=true]{hyperref}  
      
\usepackage[all]{hypcap}     

\begin{document}

\numberwithin{equation}{section}


\mathchardef\mhyphen="2D


\newcommand{\be}{\begin{equation}}
\newcommand{\ee}{\end{equation}}
\newcommand{\bea}{\begin{eqnarray}\displaystyle}
\newcommand{\eea}{\end{eqnarray}}
\newcommand{\nnm}{\nonumber}
\newcommand{\nn}{\nonumber}
\newcommand{\newotimes}{}  
\newcommand{\utv}{|0_R^{-}\rangle^{(1)}\newotimes |0_R^{-}\rangle^{(2)}}
\newcommand{\utvb}{|\bar 0_R^{-}\rangle^{(1)}\newotimes |\bar 0_R^{-}\rangle^{(2)}}

\def\eq#1{(\ref{#1})}
\newcommand{\secn}[1]{Section~\ref{#1}}

\newcommand{\tbl}[1]{Table~\ref{#1}}
\newcommand{\fig}{Fig.~\ref}

\def\beq{\begin{equation}}
\def\eeq{\end{equation}}
\def\beqa{\begin{eqnarray}}
\def\eeqa{\end{eqnarray}}
\def\bet{\begin{tabular}}
\def\eet{\end{tabular}}
\def\bs{\begin{split}}
\def\es{\end{split}}
\def\sqi{{1\over \sqrt{2}}}


\def\a{\alpha}  \def\b{\beta}   \def\c{\chi}    
\def\g{\gamma}  \def\G{\Gamma}  \def\e{\epsilon}  
\def\vep{\varepsilon}   \def\tvep{\widetilde{\varepsilon}}
\def\f{\phi}    \def\F{\Phi}  \def\fb{{\ov \phi}}
\def\vf{\varphi}  \def\m{\mu}  \def\mub{\ov \mu}
\def\n{\nu}  \def\nub{\ov \nu}  \def\o{\omega}
\def\O{\Omega}  \def\r{\rho}  \def\k{\kappa}
\def\kab{\ov \kappa}  \def\s{\sigma}
\def\t{\tau}  \def\th{\theta}  \def\sb{\ov\sigma}  \def\S{\Sigma}
\def\l{\lambda}  \def\L{\Lambda}  \def\p{\psi}

\newcommand{\gt}{\tilde{\gamma}}


\def\cA{{\cal A}} \def\cB{{\cal B}} \def\cC{{\cal C}}
\def\cD{{\cal D}} \def\cE{{\cal E}} \def\cF{{\cal F}}
\def\cG{{\cal G}} \def\cH{{\cal H}} \def\cI{{\cal I}}
\def\cJ{{\cal J}} \def\cK{{\cal K}} \def\cL{{\cal L}}
\def\cM{{\cal M}} \def\cN{{\cal N}} \def\cO{{\cal O}}
\def\cP{{\cal P}} \def\cQ{{\cal Q}} \def\cR{{\cal R}}
\def\cS{{\cal S}} \def\cT{{\cal T}} \def\cU{{\cal U}}
\def\cV{{\cal V}} \def\cW{{\cal W}} \def\cX{{\cal X}}
\def\cY{{\cal Y}} \def\cZ{{\cal Z}}

\def\mC{\mathbb{C}} \def\mP{\mathbb{P}}  
\def\mR{\mathbb{R}} \def\mZ{\mathbb{Z}} 
\def\mT{\mathbb{T}} \def\mN{\mathbb{N}}
\def\mH{\mathbb{H}} \def\mX{\mathbb{X}}

\def\CP{\mathbb{CP}}
\def\RP{\mathbb{RP}}
\def\Z{\mathbb{Z}}
\def\N{\mathbb{N}}
\def\H{\mathbb{H}}

\newcommand{\rmd}{\mathrm{d}}
\newcommand{\rmx}{\mathrm{x}}

\def\tA{ {\widetilde A} } 

\def\one{{\hbox{\kern+.5mm 1\kern-.8mm l}}}
\def\zero{{\hbox{0\kern-1.5mm 0}}}


\newcommand{\bra}[1]{{\langle {#1} |\,}}
\newcommand{\ket}[1]{{\,| {#1} \rangle}}
\newcommand{\braket}[2]{\ensuremath{\langle #1 | #2 \rangle}}
\newcommand{\Braket}[2]{\ensuremath{\langle\, #1 \,|\, #2 \,\rangle}}
\newcommand{\norm}[1]{\ensuremath{\left\| #1 \right\|}}
\def\corr#1{\left\langle \, #1 \, \right\rangle}
\def\vac{|0\rangle}


\def\d{ \partial } 
\def\zb{{\bar z}}

\newcommand{\sq}{\square}
\newcommand{\IP}[2]{\ensuremath{\langle #1 , #2 \rangle}}    

\newcommand{\floor}[1]{\left\lfloor #1 \right\rfloor}
\newcommand{\ceil}[1]{\left\lceil #1 \right\rceil}

\newcommand{\dbyd}[1]{\ensuremath{ \frac{\d}{\d {#1}}}}
\newcommand{\ddbyd}[1]{\ensuremath{ \frac{\d^2}{\d {#1}^2}}}

\newcommand{\Zd}{\ensuremath{ Z^{\dagger}}}
\newcommand{\Xd}{\ensuremath{ X^{\dagger}}}
\newcommand{\Ad}{\ensuremath{ A^{\dagger}}}
\newcommand{\Bd}{\ensuremath{ B^{\dagger}}}
\newcommand{\Ud}{\ensuremath{ U^{\dagger}}}
\newcommand{\Td}{\ensuremath{ T^{\dagger}}}

\newcommand{\T}[3]{\ensuremath{ #1{}^{#2}_{\phantom{#2} \! #3}}}		

\newcommand{\tr}{\operatorname{tr}}
\newcommand{\sech}{\operatorname{sech}}
\newcommand{\Spin}{\operatorname{Spin}}
\newcommand{\Sym}{\operatorname{Sym}}
\newcommand{\Com}{\operatorname{Com}}
\def\adj{\textrm{adj}}
\def\id{\textrm{id}}

\def\ha{\frac{1}{2}}
\def\tha{\tfrac{1}{2}}
\def\wt{\widetilde}
\def\ra{\rangle}
\def\la{\langle}

\def\pb{\ov\psi}
\def\pt{\widetilde{\psi}}
\def\at{\widetilde{\a}}
\def\cb{\ov\chi}
\def\d{\partial}
\def\db{\bar\partial}
\def\delb{\bar\partial}
\def\dbar{\ov\partial}
\def\dag{\dagger}
\def\dalpha{{\dot\alpha}}
\def\dbeta{{\dot\beta}}
\def\dgamma{{\dot\gamma}}
\def\ddelta{{\dot\delta}}
\def\ad{{\dot\alpha}}
\def\bd{{\dot\beta}}
\def\dg{{\dot\gamma}}
\def\dd{{\dot\delta}}
\def\th{\theta}
\def\Th{\Theta}
\def\eb{{\ov \epsilon}}
\def\gb{{\ov \gamma}}
\def\wb{{\ov w}}
\def\Wb{{\ov W}}
\def\D{\Delta}
\def\DD{\Delta^\dag}
\def\Db{\ov D}

\def\ov{\overline}
\def\Slash{\, / \! \! \! \!}
\def\dslash{\partial\!\!\!/} 
\def\Dslash{D\!\!\!\!/\,\,}
\def\fslash#1{\slash\!\!\!#1}
\def\Fslash#1{\slash\!\!\!\!#1}

\def\del{\partial}
\def\delb{\bar\partial}
\newcommand{\ex}[1]{{\rm e}^{#1}} 
\def\ii{{i}}

\newcommand{\vs}[1]{\vspace{#1 mm}}

\newcommand{\ve}{{\vec{\e}}}
\newcommand{\shalf}{\frac{1}{2}}

\newcommand{\lb}{\rangle}
\newcommand{\al}{\ensuremath{\alpha'}}
\newcommand{\ap}{\ensuremath{\alpha'}}

\newcommand{\bean}{\begin{eqnarray*}}
\newcommand{\eean}{\end{eqnarray*}}
\newcommand{\ft}[2]{{\textstyle {\frac{#1}{#2}} }}

\newcommand{\hsp}{\hspace{0.5cm}}
\def\half{{\textstyle{1\over2}}}
\let\ci=\cite \let\re=\ref
\let\se=\section \let\sse=\subsection \let\ssse=\subsubsection

\newcommand{\dpb}{D$p$-brane}
\newcommand{\dpbs}{D$p$-branes}

\def\gh{{\rm gh}}
\def\sgh{{\rm sgh}}
\def\NS{{\rm NS}}
\def\R{{\rm R}}
\def\Qp{Q_{\rm P}}
\def\QP{Q_{\rm P}}

\newcommand\dott[2]{#1 \! \cdot \! #2}

\def\eo{\overline{e}}


\def\p{\partial}
\def\h{{1\over 2}}

\def\d{\partial}
\def\la{\lambda}
\def\eps{\epsilon}
\def\bb{\bigskip}
\def\tg{\widetilde\gamma}
\newcommand{\dm}{\begin{displaymath}}
\newcommand{\edm}{\end{displaymath}}
\renewcommand{\b}{\widetilde{B}}
\newcommand{\gm}{\Gamma}
\newcommand{\ac}[2]{\ensuremath{\{ #1, #2 \}}}
\renewcommand{\ell}{l}
\newcommand{\z}{\ell}
\def\bb{$\bullet$}
\def\Qbar{{\bar Q}_1}
\def\QPbar{{\bar Q}_p}

\def\q{\quad}

\def\bn{B_\circ}

\let\a=\alpha \let\b=\beta \let\g=\gamma 
\let\e=\epsilon
\let\c=\chi \let\th=\theta  \let\k=\kappa
\let\l=\lambda \let\m=\mu \let\n=\nu \let\x=\xi \let\r=\rho

\let\s=\sigma

\let\vp=\varphi \let\vep=\varepsilon
\let\w=\omega  \let\G=\Gamma \let\D=\Delta \let\Th=\Theta \let\P=\Pi \let\S=\Sigma

\def\h{{1\over 2}}

\def\r{\rightarrow}
\def\Ri{\Rightarrow}

\def\nn{\nonumber\\}
\let\bm=\bibitem
\def\Kt{{\widetilde K}}
\def\b{\vspace{3mm}}
\def\t{\tilde}
\let\p=\partial


\begin{flushright}
\end{flushright}

\vspace{16mm}

 \begin{center}
{\LARGE Effect of the twist operator in the D1D5 CFT}
\\
\vspace{18mm}
{\bf  Zaq Carson\footnote{carson.231@osu.edu}, Shaun Hampton\footnote{hampton.197@osu.edu}, Samir D. Mathur\footnote{mathur.16@osu.edu} and David Turton\footnote{turton.7@osu.edu} 
\\}
\vspace{15mm}
Department of Physics,\\ The Ohio State University,\\ Columbus,
OH 43210, USA\\ 
\vspace{8mm}
\end{center}

\vspace{10mm}

\thispagestyle{empty}
\begin{abstract}

\vspace{3mm}

The D1D5 CFT has been very useful in the study of black holes. The interaction in this theory involves a twist operator, which links together different copies of a free CFT. For the bosonic fields, we examine the action of this twist when it links together CFT copies with winding numbers $M$ and $N$ to produce a copy with winding $M+N$. Starting with the vacuum state generates a squeezed state, which we compute. Starting with an initial excitation on one of the copies gives  a linear combination of excitations on the final state, which we also compute. These results generalize earlier computations where these quantities were computed for the special case $M=N=1$. Our results should help in understanding the thermalization process in the  D1D5 CFT, which gives the dual of black hole formation in the bulk.

\end{abstract}
\newpage

\section{Introduction}
\label{intr}\setcounter{footnote}{0}

\baselineskip=15pt
\parskip=3pt

String theory has had remarkable success in explaining the quantum physics of black holes. A very useful example has been the D1D5 system - a bound state of $N_1$ D1 branes and $N_5$ D5 branes. This bound state and its excitations give a dual description of black holes in 4+1 noncompact dimensions \cite{sv, adscft}. 

The dynamics of the D1D5 bound state is given by a 1+1 dimensional CFT.  The moduli space of couplings is believed to have a an `orbifold point', where the theory is essentially free \cite{orbifold}.  At this point the CFT is described by a symmetric product of $N_1N_5$ copies of a free CFT, where each copy contains 4 bosons and 4 fermions. Since the different copies are symmetrized, the operator content includes twist operators. A twist operator $\sigma_n$ takes $n$ different copies of the CFT and links them into a single copy living on a circle that is $n$ times longer. We call any such linked set of copies a `component string'. 

The `free' theory at the orbifold point has been surprisingly successful in reproducing many aspects of black hole physics. The free CFT yields exact agreement with the properties of   near-extremal black holes; for example the entropy and greybody factors are reproduced exactly \cite{sv,comparing}. 

However, the full dynamics of black holes is not given by the CFT at the orbifold point; we have to deform away from this `free' theory by an operator which corresponds to turning on the coupling constant of the orbifold theory. In particular the process of black hole formation is described in the CFT by the   thermalization of an initially non-thermal  state; such a process requires nontrivial interactions in the CFT.

\begin{figure}[t]
\begin{center}
\includegraphics[scale=.50]{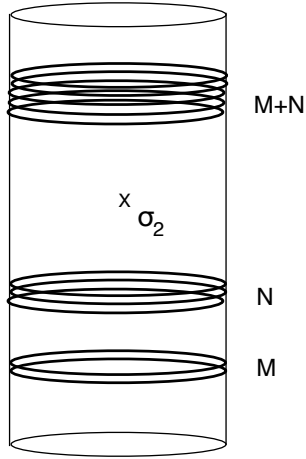}
\caption{The cylinder with coordinate $w$. The state before the twist has component strings with windings $M,N$. The twist operator $\sigma_2$  links these into a single component string of winding $M+N$.}
\label{fone}
\end{center}
\end{figure}

The deformation operator describing these interactions has the form of a twist operator $\sigma_2$, dressed with a supercharge: $\hat O\sim G_{-\h} \sigma_2$. The effect of the twist is depicted in Fig.\;\ref{fone}. Before the interaction,  we have component strings, with windings $M, N$. The interaction links these component strings together, generating a component string with winding $M+N$.\footnote{In the present paper we focus on the process of two component strings joining together. If the twist operator acts on two strands of the same component string, then it will decompose the component string into two parts. The computations for this case can be done in a similar manner to the computations presented here.}

In this paper we  study the effects of this interaction. We carry out the following computations:

(a) Firstly, we start with the vacuum state $|0^{(1)}\rangle |0^{(2)}\rangle$ on the initial component strings in Fig.\;\ref{fone}. We apply the twist operator $\sigma_2$ at a point $w_0$. This generates a component string with winding $M+N$, but this component string will not be in its vacuum state $|0\rangle$. As argued in \cite{acm2}, the starting vacuum state gets converted into a state on the final component string with the schematic form
\be
|0^{(1)}\rangle |0^{(2)}\rangle ~~\r ~~
|\chi\rangle~~\sim~~ e^{\sum_{k,l}\gamma^B_{kl}\alpha_{-k}\alpha_{-l}}|0\rangle
\ee
where $\alpha_k$ are the oscillator modes of a free scalar field.
In \cite{acm2} the coefficients $\gamma^B_{kl}$ were computed, for both bosonic and fermionic excitations, for the case where the initial component strings had windings $M=N=1$. In the present paper we will restrict attention to bosons, but find the $\gamma^B_{kl}$ for arbitrary $M,N$.

\b

(b) Secondly, we start with an initial excitation $\alpha^{(i)}_{-m}$ on one of the component strings before the twist; here $i=1,2$ labels which component string we excite. After the twist, this excitation gets converted to a linear combination of excitations above the state $|\chi\rangle$,
\be
\alpha^{(i)}_{-m} |0^{(1)}\rangle |0^{(2)}\rangle ~~\r ~~\sum\limits_k f^{B(i)}_{mk}\,  \alpha_{-k}|\chi\rangle \,, \qquad i=1,2
\ee
on the final component string of length  $M+N$. The coefficients $f^{B(i)}_{mk}$ were found for $M=N=1$ in \cite{acm3}. In the present paper, again restricting  to bosonic fields, we find these coefficients for general $M,N$. 

\b

(c) The physics of the black hole is captured by taking  $N_1N_5\gg 1$. In this limit, component strings will typically have windings $M\gg1$, and the excitations $\alpha_{-m}$ on  the component strings will typically have  have a wavelength that is much shorter than the length of the component string. Thus we will be interested in a `continuum limit' where the excitations have $m\gg 1$; such excitations are not sensitive to the infra-red cutoff scale set by the length of the component string. We find the continuum limit approximations to the expressions for $\gamma^B_{kl}, f^{B(i)}_{mk}$.

The computation of $\gamma^B_{kl}, f^{B(i)}_{mk}$ for general $M,N$ is somewhat more involved than the computation for $M=N=1$, for the following reason. For $M=N=1$ there are two twist operators: the interaction $\sigma_2$ and a twist $\sigma_2$ required to generate the component string of length $2$ in the final state. For general $M,N$, there are four twists: twists $\sigma_M, \sigma_N$ for the initial component strings, the interaction $\sigma_2$, and a twist $\sigma_{M+N}$ at infinity for the final component string.

The expressions for $\gamma^B_{kl}, f^{B(i)}_{mk}$ turn out to have a nice structure given in terms of gamma functions. The continuum limit expressions are  simpler, and may be of more use in understanding the thermalization problem.

Several other directions have been studied with the twist operator. In \cite{ac} intertwining relations were derived for operators before and after the twist. The effect of the twist on entanglement entropy was studied in \cite{aa}. Twist-nontwist correlators were computed in \cite{peet1}, and operator mixing was studied in \cite{peet2}. For other related work, see \cite{orbifold2}. Our line of enquiry complements the fuzzball program; for early work, see~\cite{fuzzball1}, for reviews see 
\cite{reviews} and for recent work, see e.g.~\cite{fuzzrecent1,fuzzrecent2}.

This paper is organized as follows. In Section \ref{sectiontwo} we introduce the D1D5 CFT and discuss the nature of the twist interaction. In Section \ref{sec:gamma} we compute the effect of the twist on the vacuum state and in Section \ref{sec:f} we compute the effect of the twist on a state with an initial excitation. In Section \ref{sec:cont} we obtain the continuum limit approximations to these quantities. In Section \ref{sec:disc} we discuss our results.

\section{The D1D5 CFT at the orbifold point}\label{sectiontwo}

In this section we summarize some properties of the D1D5 CFT at the orbifold point and the deformation operator that we will use to perturb away from the orbifold point. For more details, see \cite{acm2}.

\subsection{The D1D5 CFT}

Consider type IIB string theory, compactified as
\be
M_{9,1}\rightarrow M_{4,1}\times S^1\times T^4.
\label{compact}
\ee
Wrap $N_1$ D1 branes on $S^1$, and $N_5$ D5 branes on $S^1\times
T^4$. The bound state of these branes is described by a field
theory. We think of the $S^1$ as being large compared to the $T^4$, so
that at low energies we look for excitations only in the direction
$S^1$.  This low energy limit gives a conformal field theory (CFT) on
the circle $S^1$.

We can vary the moduli of string theory (the string coupling $g$, the
shape and size of the torus, the values of flat connections for gauge
fields etc.). These changes move us to different points in the moduli
space of the CFT. It has been conjectured that we can move to a point
called the `orbifold point' where the CFT is particularly simple
\cite{orbifold}. At this orbifold point the CFT is
a 1+1 dimensional sigma model. We will work in the Euclidean theory, where
the base space is a cylinder spanned by the coordinates 
\be
\tau, \sigma: ~~~0\le \sigma<2\pi, ~~~-\infty<\tau<\infty
\ee
The target space of the sigma model is the symmetrized product of
$N_1N_5$ copies of $T^4$,
\be
(T_4)^{N_1N_5}/S_{N_1N_5},
\ee
with each copy of $T^4$ giving 4 bosonic excitations $X^1, X^2, X^3,
X^4$. It also gives 4 fermionic excitations, which we call $\psi^1,
\psi^2, \psi^3, \psi^4$ for the left movers, and $\bar\psi^1,
\bar\psi^2,\bar\psi^3,\bar\psi^4$ for the right movers.  The central charge of the theory with fields
$X^i, \psi^i, ~i=1\dots 4$ is
$c=6$. The total central charge of the entire system is thus $6 N_1N_5$.

We will not consider the fermions in this paper; we hope to present their dynamics elsewhere. The bosons can be grouped into a matrix 
\be
X_{A\dot A}= \sqi X_i \sigma_i=\sqi\begin{pmatrix}  X_3+iX_4& X_1-iX_2\\ X_1+iX_2&-X_3+iX_4 \end{pmatrix}
\ee
where $\sigma_i=\sigma_1, \sigma_2, \sigma_3, iI$. The 2-point functions are
\be
<\p X_{A\dot A}(z) \p X_{B\dot B}(w)>~=~{1\over (z-w)^2}\epsilon_{AB}\epsilon_{\dot A\dot B}\,.
\label{ope}
\ee
In \cite{acm2} it was noted that we can write the deformation operator as
\be
\hat O_{\dot A\dot B}(w_0)=\Big [{1\over 2\pi i} \int _{w_0} dw G^-_{\dot A} (w)\Big ]\Big [{1\over 2\pi i} \int _{\bar w_0} d\bar w \bar G^-_{\dot B} (\bar w)\Big ]\sigma_2^{++}(w_0)
\label{pert}
\ee
Since we do not consider fermions in the present paper, we ignore the spins on the twist operator and also the action of the supercharge $G$. That is, for the purposes of our computations, the deformation operator will be simply 
$\sigma_2$. Throughout the paper we shall write expressions for left-movers only; analogous expressions hold for right-movers.

\subsection{Nature of the twist interaction}

\begin{figure}[t]
\begin{center}
\includegraphics[scale=.55]{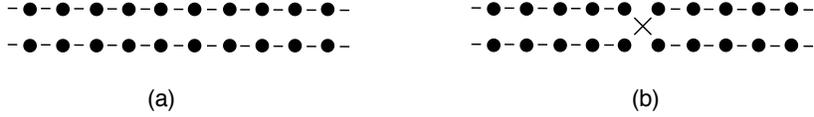}
\caption{(a) The scalar field on the component strings modeled by point masses joined by springs. (b) The twist operator $\sigma_2$ changes the springs so that the masses are linked in a different way.}
\label{ftwo}
\end{center}
\end{figure}

To understand the effect of such a twist $\sigma_2$, consider a discretization of a 1+1 dimensional bosonic free field $X$.  We can model this field by a collection of point masses joined by springs. This gives a set of coupled harmonic oscillators, and the oscillation amplitude of the masses then gives the field $X(\tau,  \sigma)$. Consider such a collection of point masses on two different circles, and let the state in each case be the ground state of the coupled oscillators (Fig.\;\ref{ftwo}(a)). At time $\tau_0$ and position $\sigma_0$, we insert a twist $\sigma_2$. The effect of this twist is to connect the masses with a different set of springs, so that the masses make a single chain of longer  length (Fig.\;\ref{ftwo}(b)).

To see what we might expect of such an interaction, consider a single harmonic oscillator,  which starts in its ground state. The Hamiltonian can be expressed in terms of annihilation and creation operators $\hat a,  {\hat a}^\dagger$, and the ground state  $|0\rangle_a$ is given by $\hat a|0\rangle_a=0$.  At time $\tau=\tau_0$, imagine changing the spring constant to a different value. The wavefunction does not change at this instant, and the Lagrangian remains quadratic. But the ground state with this new spring constant is a different state $|0\rangle_b$, and the operator $\hat {a}$ can be expressed as a linear combination of the new annihilation and creation operators $\hat b, {\hat b}^\dagger$:
\be
\hat a = \alpha \hat b +\beta {\hat b}^\dagger
\ee
The wavefunction after this change of coupling can be reexpressed as
\be
|0\rangle_a=Ce^{-\h \gamma {\hat b}^\dagger {\hat b}^\dagger}|0\rangle_b
\label{exp}
\ee
where $\g=\a^{-1} \beta$ and $C$ is a constant.
If we had a single initial excitation before the twist, it will give a single excitation after the twist, but with a nontrivial coefficient\footnote{We have $f=\alpha^*-\beta^*\gamma$.} $f$
\be
\hat a ^\dagger |0\rangle_a=f \, {\hat b}^\dagger \, Ce^{-\h \gamma {\hat b}^\dagger {\hat b}^\dagger}|0\rangle_b \,.
\label{ff}
\ee

Now let us return to the CFT. Regarding the scalar field on the component strings as a set of coupled harmonic oscillators, we note that the twist interaction changes the coupling matrix between the oscillators but not the wavefunction itself. Thus the effect of the twist is captured by reexpressing the state before the twist in terms of the natural oscillators after the twist,
\be
\alpha_{A\dot A, k}= {1\over 2\pi} \int\limits_{\sigma=0}^{2\pi (M+N)} \p_w X_{A\dot A}(w) e^{{k\over M+N}w} dw \,.
\label{qaone-0}
\ee
 In analogy to (\ref{exp})  the state
 after the twist will then have the form
\bea
|\chi\rangle&\equiv& \sigma_2(w_0)|0^{(1)}\rangle\otimes |0^{(2)}\rangle\nn
&=&C(w_0)e^{\sum_{ k\ge 1, l\ge 1}\gamma^B_{kl}(-\alpha_{++, -k}\alpha_{--, -l}+\alpha_{-+, -k}\alpha_{+-, -l})}
|0\rangle \,.
\label{pfive}
\eea
The index structure on the $\alpha$ oscillators is arranged to obtain a singlet under the group of rotations in the $T^4$. This is explained in detail in  \cite{acm2}, where we also fix 
the normalization $C(w_0)$  to unity after we include the fermions \cite{acm2}. Since we do not consider fermions in this paper, we do not seek to determine $C(w_0)$; our goal is to find the $\gamma^B_{kl}(w_0)$. 

An initial excitation on one of the component strings will transform to a linear combination of excitations on the final component string above the state $|\chi\rangle$. Analogous to (\ref{ff}), we will have
\bea
&&\sigma_2(w_0)\alpha^{(i)}_{A\dot A, -m}|0\rangle^{(1)}\otimes |0\rangle^{(2)}=\nn
&&\qquad\qquad\sum_{k\ge 1}~f^{B(i)}_{mk}~\alpha_{A\dot A, -k} ~
 e^{\sum_{ k'\ge 1, l'\ge 1}\gamma^B_{k'l'}(-\alpha_{++, -k'}\alpha_{--, -l'}+\alpha_{-+, -k'}\alpha_{+-, -l'})}|0\rangle \,.\qquad\quad
 \label{wfourqq}
\eea
where $i=1,2$ for the initial component strings with windings $M,N$ respectively. We will find the $f^{B(i)}_{mk}$.

\section{Effect of the twist operator on the vacuum state}\label{sec:gamma}

Consider the process depicted in Fig.\;\ref{fone}. We insert the twist operator $\sigma_2$ at a location 
\be
w_0=\tau_0+i\sigma_0
\ee
The twist operator changes the Lagrangian of the theory from one free Lagrangian (that describes  free CFTs on circles of length $2\pi M$ and $2\pi N$) to another free Lagrangian (that for a free CFT on a circle of length $2\pi(M+N)$). As explained in \cite{acm2}, in this situation the vacuum state of the initial theory does not go over to the vacuum of the new theory. But the excitations must take the special form of a Gaussian (\ref{pfive}), so our goal is to find the coefficients $\gamma^B_{kl}$. 

The steps we follow are analogous to those in \cite{acm2}. We first map the cylinder $w$ to the complex plane through $z=e^w$. The CFT field $X$ will be multivalued in the $z$ plane, due to the presence of twist operators.  The initial component strings with  windings $M,N$ are created by twist operators $\sigma_M, \sigma_N$. The interaction is another twist operator $\sigma_2$. The point at infinity has a twist of order $M+N$ corresponding to the component string in the final state.

To handle the twist operators, we go to a covering space $t$ where $X$ is single valued. The twist operators become simple punctures in the $t$ plane, with no insertions at these punctures. We can therefore trivially close these punctures. The nontrivial physics is now encoded in the definition of oscillator modes on the $t$ plane -- the creation operators on the cylinder are linear combinations of creation and annihilation operators on the $t$ plane.  Performing appropriate Wick contractions, we obtain the $\gamma^B_{kl}$. Finally, we change notation to a form that will be more useful in the situation where $k\gg 1, l\gg 1$.

\subsection{Modes on the cylinder $w$}

Let us begin by defining operator modes on the cylinder.
Below the twist insertion ($\tau<\tau_0$)
we  have modes $\alpha^{(1)}_{A\dot A, m}$ on the component string of winding $M$ and modes $\alpha^{(2)}_{A\dot A, n}$ on the component string of winding $N$:
\be
\alpha^{(1)}_{A\dot A, m}= {1\over 2\pi} \int\limits_{\sigma=0}^{2\pi M} \p_w X^{(1)}_{A\dot A}(w) e^{{m\over M}w} dw
\ee
\be
\alpha^{(2)}_{A\dot A, n}= {1\over 2\pi} \int\limits_{\sigma=0}^{2\pi N} \p_w X^{(2)}_{A\dot A}(w) e^{{n\over N}w} dw
\ee
From (\ref{ope}), we find that the commutation relations are
\be
[\alpha^{(i)}_{A\dot A, m}, \alpha^{(j)}_{B\dot B, n}]=-\epsilon_{AB}\epsilon_{\dot A\dot B}\delta^{ij} m \delta_{m+n,0}
\ee

Above this twist insertion ($\tau>\tau_0$) we have a single component string of winding $M+N$. The modes are
\be
\alpha_{A\dot A, k}= {1\over 2\pi} \int\limits_{\sigma=0}^{2\pi (M+N)} \p_w X_{A\dot A}(w) e^{{k\over M+N}w} dw
\label{qaone}
\ee
The commutation relations are
\be\label{bcommtwist}
[\alpha_{A\dot A, k}, \alpha_{B\dot B, l}]=-\epsilon_{AB}\epsilon_{\dot A\dot B}\,  k\,  \delta_{k+l,0}\;.
\ee

\subsection{Modes on the $z$ plane}

We wish to go to a covering space where the field $X_{A\dot A}$ will be single valued.  As a preparatory step, it is convenient to map the cylinder with coordinate $w$ to the plane with coordinate $z$,
\be
z=e^w
\ee
Under this map the operator modes change as follows. Before the insertion of the twist ($|z|<e^{\tau_0}$) we have,  using a contour circling ${z=0}$
\be
\alpha^{(1)}_{A\dot A, m}\r {1\over 2\pi} \int\limits_{{\rm arg}(z)=0}^{2\pi M} \p_z X^{(1)}_{A\dot A}(z) z^{m\over M} dz
\ee
\be
\alpha^{(2)}_{A\dot A, n}\r {1\over 2\pi} \int\limits_{{\rm arg}(z)=0}^{2\pi N} \p_z X^{(2)}_{A\dot A}(z) z^{n\over N} dz
\ee

After the twist ($|z|>e^{\tau_0}$) we have, using a contour circling ${z=\infty}$
\be
\alpha_{A\dot A, k}\r {1\over 2\pi} \int\limits_{{\rm arg}(z)=0}^{2\pi (M+N)} \p_z X_{A\dot A}(z) z^{{k\over M+N}} dz \,.
\ee

\subsection{Modes on the covering space $t$}

We now proceed to the covering space $t$ where $X_{A\dot A}$ will be single-valued. Consider the map
\be
z=t^M(t-a)^N
\label{cover}
\ee
The various operator insertions map as follows:

\b

(i) The initial component strings were at $w\r-\infty$ on the cylinder, which corresponds to $z=0$ on the $z$ plane. In the $t$ plane, the component string of winding $M$ maps to $t=0$, while the component string with winding $N$ maps to $t=a$.

(ii) The final component string state is at $w\r\infty$ on the cylinder, which corresponds to $z=\infty$. This maps to $t=\infty$.

(iii) The twist operator $\sigma_2$ is at $w_0$ on the cylinder, which corresponds to $e^{w_0}$ on the $z$ plane. To find its location on the covering space $t$, we note that ${dz\over dt}$ should vanish at the location of every twist, since these are ramification points of map (\ref{cover}) to the covering space. We find that apart from the ramification points at $t=0,a, \infty$, the function  ${dz\over dt}$ also vanishes at
\be
t={a M\over M+N}
\ee
which corresponds to the following value of $z$:
\be
z_0={a^{M+N}} {M^MN^N\over (M+N)^{M+N}}(-1)^N \,.
\ee
To solve for $a$, we must specify how to deal with the fractional exponent. We do this by choosing
\be
z_0={a^{M+N}} {M^MN^N\over (M+N)^{M+N}}e^{i\pi N} \,
\ee
which determines the quantity $a$ in terms of $z_0=e^{w_0}$, as
\be
a=e^{-i\pi \frac{N}{M+N}}\left( {z_0\over M^MN^N}\right)^{1\over M+N}(M+N)\,.
\label{az}
\ee

Now let us consider the modes in the $t$ plane. Before the twist we have
\be
\alpha^{(1)}_{A\dot A, m}\r {1\over 2\pi} \oint\limits_{t=0} dt \, \p_t X_{A\dot A}(t)  \left ( t^M(t-a)^N \right ) ^{m\over M}
\ee
\be
\alpha^{(2)}_{A\dot A, n}\r {1\over 2\pi} \oint\limits_{t=a} dt \, \p_t X_{A\dot A}(t)  \left ( t^M(t-a)^N \right ) ^{n\over N}
\ee
After the twist we have
\be
\alpha_{A\dot A, k}\r {1\over 2\pi}\oint\limits_{t=\infty} dt \, \p_t X_{A\dot A}(t)  \left ( t^M(t-a)^N \right ) ^{k\over M+N}
\ee
We also define mode operators that are natural to the $t$ plane, as follows:
\be
\t\alpha_{A\dot A, m}\equiv  {1\over 2\pi} \oint\limits_{t=0} dt \, \p_t X_{A\dot A}(t) t^m 
\label{qathree}
\ee
The commutation relations are
\be
[\t\alpha_{A \dot A, k}, \t\alpha_{B \dot B, l}]=-\epsilon_{A B} \epsilon_{\dot A \dot B} \, k\, \delta_{k+l, 0}
\label{pttwo}
\ee

\subsection{Method for finding  the  $\gamma^B_{kl}$}

Let us consider the amplitude
\bea
{\cal A}_1&=&\langle 0|\sigma_2(w_0)|0\rangle^{(1)}\otimes |0\rangle^{(2)}\nn
&=&C(w_0)\langle 0|e^{\sum_{ k\ge 1, l\ge 1}\gamma^B_{kl}[-\alpha_{++, -k}\alpha_{--, -l}+\alpha_{-+, -k}\alpha_{+-, -l}]}
|0\rangle\nn
&=& C(w_0)
\eea
where we assume that the vacuum is normalized as $\langle 0 | 0 \rangle=1$.

We compare this to the amplitude
\bea
{\cal A}_2&=&\langle 0|\Big (\alpha_{++,_l}\alpha_{--,_k}\Big )\sigma_2(w_0)|0\rangle^{(1)}\otimes |0\rangle^{(2)}\nn
&=&C(w_0)\langle 0|\Big (\alpha_{++,_l}\alpha_{--,_k}\Big )e^{\sum_{ k'\ge 1, l'\ge 1}\gamma^B_{k'l'}[-\alpha_{++, -k'}\alpha_{--, -l'}+\alpha_{-+, -k'}\alpha_{+-, -l'}]}
|0\rangle\nn
&=& -C(w_0)~kl\gamma^B_{kl}~\langle 0|0\rangle=-C(w_0)~kl\gamma^B_{kl}
\eea
Thus we see that
\be
\gamma^B_{kl}=-{1\over kl} {{\cal A}_2\over {\cal A}_1} \,.
\ee

To compute ${\cal A}_1$ we map the cylinder $w$ to the plane $z$ and then to the cover $t$. In this cover the locations of the twist operators are just punctures with no insertions. Thus these punctures can be closed, making the $t$ space just a sphere. Closing the punctures involves normalization factors, so we write
\be
{\cal A}_1=\langle 0|\sigma_2(w_0)|0\rangle^{(1)}\otimes |0\rangle^{(2)}=D(z_0)~{}_t\langle 0 |0\rangle_t
\ee
Factors like $D(z_0)$ were computed in \cite{lm1}, but here we do not need to compute $D(z_0)$ since it will cancel in the ratio ${\cal A}_2/{\cal A}_1$. We have
\be
{\cal A}_2=\langle 0|\Big (\alpha_{++,_l}\alpha_{--,_k}\Big )\sigma_2(w_0)|0\rangle^{(1)}\otimes |0\rangle^{(2)}=D(z_0)~{}_t\langle 0| \Big (\alpha'_{++,_l}\alpha'_{--,_k}\Big )|0\rangle_t
\ee
where the primes on the operators on the RHS signify the fact that these operators arise from the unprimed operators by the various maps leading to the $t$ plane description. Thus we have
\be
\gamma^B_{kl}=-{1\over kl} {{\cal A}_2\over {\cal A}_1}=-{1\over kl} ~{{}_t\langle 0| \Big (\alpha'_{++,_l}\alpha'_{--,_k}\Big )|0\rangle_t\over {}_t\langle 0 |0\rangle_t}\,.
\ee

\subsection{Computing the $\gamma^B_{kl}$}

The operators $\alpha'$ are given by contour integrals at large $t$:
\bea
&&\hskip-.25in{}_t\langle 0| \Big (\alpha'_{++,_l}\alpha'_{--,_k}\Big )|0\rangle_t=\nn
&&\hskip-.25in{}_t\langle 0 |\Big ({1\over 2\pi }\int_{} dt_1 \p_t X_{++}(t_1) \left ( t_1^M(t_1-a)^N \right ) ^{l\over M+N}\Big )\Big(
{1\over 2\pi }\int_{} dt_2 \p_t X_{--}(t_2) \left ( t_2^M(t_2-a)^N \right ) ^{k\over M+N}\Big )
|0\rangle_t\nn
\label{ptfive}
\eea
with $|t_1|>|t_2|$. We have\footnote{The symbol ${}^nC_m$ is the binomial coefficient, also written $\begin{pmatrix} n\\ m\\ \end{pmatrix}$.}
\be
\left ( t_1^M(t_1-a)^N \right ) ^{l\over M+N}=t_1^l(1-a t_1^{-1})^{lN\over M+N}=t_1^l\sum_{p\ge 0} {}^{lN\over M+N}C_p (-a)^p t_1^{-p}=\sum_{p\ge 0} {}^{lN\over M+N}C_p (-a)^p t_1^{l-p}
\ee
\be
\left ( t_2^M(t_2-a)^N \right ) ^{k\over M+N}=t_2^k(1-a t_2^{-1})^{kN\over M+N}=t_2^k\sum_{q\ge 0} {}^{kN\over M+N}C_q (-a)^q t_2^{-q}=\sum_{q\ge 0} {}^{kN\over M+N}C_q (-a)^q t_2^{k-q}
\ee
Thus
\bea
{1\over 2\pi }\int_{} dt_1 \p_t X_{++}(t_1) \left ( t_1^M(t_1-a)^N \right ) ^{l\over M+N}&=&\sum_{p\ge 0} {}^{lN\over M+N}C_p (-a)^p \t\alpha_{++, {l-p}}\nn
{1\over 2\pi }\int_{} dt_2 \p_t X_{--}(t_2) \left ( t_2^M(t_2-a)^N \right ) ^{k\over M+N}&=&\sum_{q\ge 0} {}^{kN\over M+N}C_q (-a)^q \t\alpha_{--, {k-q}}
\eea
We then find
\be
\gamma^B_{kl}=-{1\over kl}\sum_{p\ge 0} \sum_{q\ge 0}~{}^{lN\over M+N}C_p ~{}^{kN\over M+N}C_q (-a)^{p+q}~ {{}_t\langle 0|\t \alpha_{++,l-p}~ \t \alpha_{--,k-q}|0\rangle_t\over {}_t\langle 0|0\rangle_t}
\ee
Using the commutation relations (\ref{pttwo}) we get
\be
k-q=-(l-p) ~\Rightarrow ~q=(k+l)-p
\ee
This gives
\be
\gamma^B_{kl}={(-a)^{k+l}\over kl}\sum_{p\ge 0} ~{}^{lN\over M+N}C_p ~{}^{kN\over M+N}C_{(k+l)-p} ~(l-p)
\ee
Note that in order to give a non-zero contribution, $\t\alpha_{++,l-p}$ needs to be an annihilation operator, so we require $p\le l$. Thus we have
\be
\gamma^B_{kl}={(-a)^{k+l}\over kl}\sum_{p= 0}^l ~{}^{lN\over M+N}C_p ~{}^{kN\over M+N}C_{(k+l)-p} ~(l-p)
\ee
Evaluating this sum gives
\be
\gamma^B_{kl}=-{(-a)^{k+l}\sin[{\pi  Mk\over M+N}]\sin[{\pi Ml\over M+N}]\over \pi^2}{MN\over (M+N)^2}{1\over (k+l)}{\Gamma[{Mk\over M+N}]\Gamma[{Nk\over M+N}]\over \Gamma[k]}{\Gamma[{Ml\over M+N}]\Gamma[{Nl\over M+N}]\over \Gamma[l]}\,.
\label{gammapre}
\ee

\subsection{Expressing $\gamma^B_{kl}$ in final form}

 It will be convenient to write the expression for $\gamma^B_{kl}$ in a slightly different notation. We make the following changes:
 
 \b
 
 (i) We can replace the parameter $a$ by the variable $z_0=e^{ w_0}$ through (\ref{az}).

(ii) In addressing the continuum limit it is useful to use fractional mode numbers  defined as
\be
s={k\over M+N}, ~~~s'={l\over M+N}
\ee
The parameters $s,s'$ directly give the physical wavenumbers of the modes on the cylinder with coordinate $w$. 
When using $s,s'$ in place of $k,l$, we will write
\be
\gamma^B_{kl}\r \t\gamma^B_{ss'}\,.
\ee

(iii) We define the useful shorthand notation
\be
1-e^{2\pi i Ms}=\mu_s \,.
\ee
Note that
\be
(-1)^k\sin[{\pi  Mk\over M+N}]= (-1)^{(M+N)s}\sin(\pi Ms)={i\over 2}(-1)^{Ns}(1-e^{2\pi i Ms})=  {i\over 2}(-1)^{Ns}\mu_s \,.
\ee

\b

With these changes of notation we find
\bea
\t\gamma^B_{ss'}&=&{1\over 4\pi^2}\,z_0^{s+s'}\,
{\mu_s\mu_{s'}\over  s+s'}\,{MN\over (M+N)^3} 
\left ({(M+N)^{M+N}\over M^MN^N}\right )^{s+s'} 
{\Gamma[{Ms}]\Gamma[{Ns}]\over \Gamma[(M+N)s]}~{\Gamma[{Ms'}]\Gamma[{Ns'}]\over \Gamma[(M+N)s']}\,.\nn
\label{gammafinal}
\eea

\subsection{The case $M=N=1$}

In \cite{acm2} the $\gamma^B_{ss'}$ were computed for the case $M=N=1$. Let us check that our general result reduces to the result in \cite{acm2} for these parameters.

Our general expression for $\gamma^B_{ss'}$ is given in (\ref{gammafinal}). 
For $M=N=1$ we have
\be
\Gamma[(M+N) s]=\Gamma[2s]={2^{2s-\h}\over (2\pi)^\h} \Gamma[s]\Gamma[s+\h]
\label{gammahalf}
\ee
Substituting in (\ref{gammafinal}) we get
\bea
\t\gamma^B_{ss'}={z_0^{s+s'}\over 2\pi (s+s')}\,
{\Gamma[{s}]\over \Gamma[s+\h]}\,
{\Gamma[{s'}]\over \Gamma[s'+\h]}
\eea
which agrees with the result in \cite{acm2}. 

\section{Effect of the twist operator on an initial excitation}\label{sec:f}

We now turn to the case where one of the initial component strings has an oscillator excitation. We denote this excitation by $\alpha^{(1)}_{A\dot A, -m}$ for the component string with winding $M$ and by $\alpha^{(2)}_{A\dot A, -m}$ for the component string with winding $N$. 
We write
\bea
&&\sigma_2(w_0)\alpha^{(1)}_{A\dot A, -m}|0\rangle^{(1)}\otimes |0\rangle^{(2)}=\nn
&&\qquad\qquad\sum_{k\ge 1}~f^{B(1)}_{mk}~\alpha_{A\dot A, -k} ~
 e^{\sum_{ k'\ge 1, l'\ge 1}\gamma^B_{k'l'}(-\alpha_{++, -k'}\alpha_{--, -l'}+\alpha_{-+, -k'}\alpha_{+-, -l'})}|0\rangle \qquad
 \label{wfour}
\eea
\bea
&&\sigma_2(w_0)\alpha^{(2)}_{A\dot A, -m}|0\rangle^{(1)}\otimes |0\rangle^{(2)}=\nn
&&\qquad\qquad\sum_{k\ge 1}~f^{B(2)}_{mk}~\alpha_{A\dot A, -k} ~
 e^{\sum_{ k'\ge 1, l'\ge 1}\gamma^B_{k'l'}(-\alpha_{++, -k'}\alpha_{--, -l'}+\alpha_{-+, -k'}\alpha_{+-, -l'})}|0\rangle \qquad
 \label{wfourq}
\eea
In this section we find the functions $f^{B(i)}_{mk}$.

\subsection{Method for finding the $f^{B(i)}_{mk}$}

Analogously to the computation of $\gamma^B_{kl}$, let us consider the amplitude
\bea
{\cal A}_3&=&\langle 0|\alpha_{--,_k}\, \sigma_2(w_0)\, \alpha^{(1)}_{++, -m}|0\rangle^{(1)}\otimes |0\rangle^{(2)}\nn
&=&C(w_0)\sum_{l\ge 1}~f^{B(1)}_{ml}~\langle 0|\alpha_{--,_k}\, ~\alpha_{++, -l} ~
 e^{\sum_{ k'\ge 1, l'\ge 1}\gamma^B_{k'l'}(-\alpha_{++, -k'}\alpha_{--, -l'}+\alpha_{-+, -k'}\alpha_{+-, -l'})}|0\rangle\nn
  &=& C(w_0)\sum_{l\ge 1}~f^{B(1)}_{ml} \, (-k) \delta_{k l} \nn
&=& -C(w_0)\, k\, f^{B(1)}_{mk}\,.
\eea
In the second step above, we note that there is also a contribution when $\alpha_{--,_k}$ contracts with the terms in the exponential, but this contribution  consists of $\alpha_{++, -k'}$ with $k'>0$. Such oscillators annihilate the vacuum $\langle 0|$, and so this contribution in fact vanishes. 

Thus we see that
\be
f^{B(1)}_{mk}=-{1\over k} {{\cal A}_3\over {\cal A}_1}\,.
\ee
We have
\be
{\cal A}_3=\langle 0|\alpha_{--,_k}\sigma_2(w_0)\alpha^{(1)}_{++, -m}|0\rangle^{(1)}\otimes |0\rangle^{(2)}=D(z_0)~{}_t\langle 0| \alpha'_{--,_k}\alpha'^{(1)}_{++, -m}|0\rangle_t
\ee
Thus we obtain
\be
f^{B(1)}_{mk}=-{1\over k} {{\cal A}_3\over {\cal A}_1}=-{1\over k}~{{}_t\langle 0| \alpha'_{--,_k}\alpha'^{(1)}_{++, -m}|0\rangle_t\over {}_t\langle 0 |0\rangle_t}\,.
\label{wsix}
\ee

\subsection{Computing  $f^{B(1)}_{mk}$}

Let us now carry out the details of the  computation we outlined above. 

The operator $\alpha'_{--,k}$ is applied at $w=\infty$, and is thus given in the $t$ plane by a contour at large $t$. The operator $\alpha'^{(1)}_{++, -m}$ on the other hand is applied at $w=-\infty$ to the component string with winding $M$, and is thus given in the $t$ plane by a contour around $t=0$. Thus we get
\bea
&&{}_t\langle 0|\alpha'_{--,_k}\alpha'^{(1)}_{++, -m}|0\rangle_t=
{}_t\langle 0 |\Big ({1\over 2\pi }\int\limits_{t_1=\infty} dt_1 \p_t X_{--}(t_1) \left ( t_1^M(t_1-a)^N \right ) ^{k\over M+N}\Big )
\nn
&&\qquad\qquad \qquad\qquad\qquad\quad\times ~~
\Big(
{1\over 2\pi }\int\limits_{t_2=0} dt_2 \p_t X_{++}(t_2) \left ( t_2^M(t_2-a)^N \right ) ^{-{m\over M}}\Big )
|0\rangle_t  \;. \qquad\quad
\label{ptfiveq}
\eea
Since $t_1$ is large, we expand as
\be
\left ( t_1^M(t_1-a)^N \right ) ^{k\over M+N}=\sum_{p'\ge 0} {}^{Nk\over M+N} C_{p'} (-a)^{p'} t_1^{k-p'}
\ee
Thus we find
\be
\alpha'_{A\dot A, k}=\sum_{p'\ge 0} {}^{Nk\over M+N} C_{p'} (-a)^{p'}\t\alpha_{A\dot A, k-p'}\;.
\ee
Next, since $t_2\approx 0$  we  expand as
\bea
\left ( t_2^M(t_2-a)^N \right ) ^{-{m\over M}}&=&t_2^{-m}(t_2-a)^{-{mN\over M}}=t_2^{-m}(-a)^{-{mN\over M}}(1-{t_2\over a})^{-{mN\over M}}\nn
&=&t_2^{-m}(-a)^{-{mN\over M}}\sum_{p\ge 0} {}^{-{mN\over M}}C_p\, (-a)^{-p}t_2^{p}=\sum_{p\ge 0} {}^{-{mN\over M}}C_p\, (-a)^{-{mN\over M}-p}t_2^{p-m}\nn
\eea
Thus we find
\be
\alpha'^{(1)}_{A\dot A, -m}= \sum_{p\ge 0} {}^{-{mN\over M}}C_p\, (-a)^{-{mN\over M}-p}~ \t\alpha_{A\dot A, p-m}
\ee
Since $\t\alpha_{A\dot A, p-m}|0\rangle_t=0$ for $p-m\ge 0$, we get a contribution only from  $p<m$ in the above sum. Thus we have
\be
\alpha'^{(1)}_{A\dot A, -m}= \sum_{p= 0}^{m-1} {}^{-{mN\over M}}C_p\, (-a)^{-{mN\over M}-p}~ \t\alpha_{A\dot A, p-m}
\label{wone}
\ee
Using these expansions, we obtain
\bea
&&{}_t\langle 0 |\alpha'_{--,_k}\alpha'^{(1)}_{++, -m}|0\rangle_t\nn
&&=\sum_{p= 0}^{m-1} \sum_{p'\ge 0}~{}^{-{mN\over M}}C_p\,{}^{Nk\over M+N} C_{p'}~ (-a)^{-{mN\over M}-p+p'}~{}_t\langle 0 |\, \t\alpha_{--, k-p'}\, \t\alpha_{++, p-m}\,  |0\rangle_t\nn
&&=\sum_{p= 0}^{m-1} \sum_{p'\ge 0}~{}^{-{mN\over M}}C_p\,{}^{Nk\over M+N} C_{p'}~ (-a)^{-{mN\over M}-p+p'}~(p'-k)\delta_{k-p'+p-m,0}~{}_t\langle 0|0\rangle_t\nn
&&=(-a)^{k-{(M+N)m\over M}}\sum_{p= {\max(m-k, 0)}}^{m-1}~{}^{-{mN\over M}}C_p\,{}^{Nk\over M+N} C_{k+p-m} (p-m)~{}_t\langle 0|0\rangle_t\nn
&&=-{(-1)^m k\sin(\pi {Mk\over M+N})\over \pi (M+N)}\;
{(-a)^{k-{(M+N)m\over M}}\over {k\over M+N}-{m\over M}}\;
{\Gamma[{(M+N) m\over M}]\over \Gamma[m]\Gamma[{Nm\over M}]}\,
{\Gamma[{Nk\over M+N}]\Gamma[{Mk\over M+N}]\over \Gamma[k]}
~{}_t\langle 0|0\rangle_t\nn
\label{wseven}
\eea
and using (\ref{wsix}), we obtain
\be
f^{B(1)}_{mk}~=~{(-1)^m \sin(\pi {Mk\over M+N})\over \pi (M+N)}\;
{(-a)^{k-{(M+N)m\over M}}\over {k\over M+N}-{m\over M}}\;
{\Gamma[{(M+N) m\over M}]\over \Gamma[m]\Gamma[{Nm\over M}]}\,
{\Gamma[{Nk\over M+N}]\Gamma[{Mk\over M+N}]\over \Gamma[k]}
\;.
\label{fpre}
\ee
Note that in the above expression, when we have 
\be
{k\over M+N}={m\over M} \,,
\ee
both numerator and denominator vanish, since
\be
\sin(\pi {Mk\over M+N}) ~=~ \sin(\pi {m}) ~=~ 0 \,.
\ee
Since the above expression for $f^{B(1)}_{mk}$ is indeterminate in this situation, we return to the sum in the third  line of (\ref{wseven}), and take parameter values 
 \be
 m=M c, ~~~k=(M+N) c \,.
 \ee
 Here $c={j\over Y}$, where $j$ is a positive integer and $Y=\gcd(M,N)$. Then we have
 \bea
 {}_t\langle 0 |\alpha'_{--,_k}\alpha'^{(1)}_{++, -m}|0\rangle_t&=&(-a)^{-{(M+N)m\over M}+k}\sum_{p= {\max(m-k, 0)}}^{m-1}~{}^{-{mN\over M}}C_p\,{}^{Nk\over M+N} C_{k+p-m} (p-m)\nn
 &=&\sum_{p=0}^{m-1}~{}^{-{Nc}}C_p\,{}^{Nc} C_{Nc+p} (p-Mc)\,.
 \eea
 We note that since $Nc$ is a positive integer,
 \be
 {}^{Nc} C_{Nc+p}=0, ~~p>0
 \ee
 Thus only the $p=0$ term survives in the above sum, and we get
 \be
 {}_t\langle 0 |\alpha'_{--,_k}\alpha'^{(1)}_{++, -m}|0\rangle_t=-Mc
 \ee
 Thus
 \be
 f^{B(1)}_{mk}={Mc\over k}={Mc\over (M+N)c}={M\over M+N} \,.
 \ee
So for this special case, we write
\be \label{fpre2}
f^{B(1)}_{mk}|_{{k\over M+N}={m\over M}}~=~{M\over M+N} \,.
\ee

Using the gamma function identity (\ref{gammahalf}), one can check that the results (\ref{fpre}), (\ref{fpre2}) agree with the corresponding expression obtained in \cite{acm3} for the case $M=N=1$. 
 
\subsection{Expressing $f^{B(1)}_{mk}$ in final form}

We make the same changes of notation that we did for the $\gamma^B_{lk}$\;:

\b

(i) We replace $a$ by $z_0$.

(ii) We use fractional modes
\be
q={m\over M}, ~~~s={k\over M+N} \,.
\ee
 When expressing our result using $q$ and $s$ in place of $m$ and $k$, we will write
 \be
 f^{B(1)}_{mj}\r \t f^{B(1)}_{qs}
 \ee

(iii) We again use the shorthand $ \mu_s=(1-e^{2\pi i Ms})$. In doing this we note that
 \be
(-1)^m\sin(\pi  {Mk\over M+N})= (-1)^{Mq}\sin(\pi M s)={i\over 2}e^{-i\pi M(s-q)}(1-e^{2\pi i Ms}) \,.
 \ee
With these changes of notation we get, for $s \ne q$,
 \bea
\t f^{B(1)}_{qs}&=&
{i\over 2\pi}
z_0^{s-q}
{\mu_s\over s-q}\,
{1\over (M+N)}\left( {(M+N)^{M+N}\over M^MN^N}\right)^{(s-q)}
{\Gamma[(M+N)q]\over \Gamma[Mq]\Gamma[Nq]}\,
 {\Gamma[Ms]\Gamma[Ns]\over \Gamma[(M+N)s]} \nn
 \label{exactft}
 \eea
 and for $s =q$ we have
 \be
 \t f^{B(1)}_{qs}|_{q=s}={M\over M+N}\,.
 \ee

\subsection{Computing  $f^{B(2)}_{mk}$}

The expression for $f^{B(2)}_{mk}$ can be obtained by a similar computation. There are only a few changes, which we mention here.
In place of (\ref{ptfiveq}) we have
\bea
&&{}_t\langle 0|\alpha'_{--,_k}\alpha'^{(2)}_{++, -m}|0\rangle_t=
{}_t\langle 0 |\Big ({1\over 2\pi }\int\limits_{t_1=\infty} dt_1 \p_t X_{--}(t_1) \left ( t_1^M(t_1-a)^N \right ) ^{k\over M+N}\Big )
\nn
&&\qquad\qquad \qquad\qquad\qquad\quad\times ~~
\Big(
{1\over 2\pi }\int\limits_{t_2=a} dt_2 \p_t X_{++}(t_2) \left ( t_2^M(t_2-a)^N \right ) ^{-{m\over N}}\Big )
|0\rangle_t\qquad\quad
\label{ptfiveqq}
\eea
The power ${m\over M}$ has been replaced by ${m\over N}$, and the $t_2$ contour is now around $t_2=a$ instead of around $t_2=0$. We define a shifted coordinate in the $t$ plane
\be
t'=t-a
\ee
which gives
\bea
&&{}_t\langle 0|\alpha'_{--,_k}\alpha'^{(2)}_{++, -m}|0\rangle_t=
{}_t\langle 0 |\Big ({1\over 2\pi }\int\limits_{t'_1=\infty} dt'_1 \p_t X_{--}(t'_1) \left ( {t'_1}^N(t'_1+a)^M \right ) ^{k\over M+N}\Big )
\nn
&&\qquad\qquad \qquad\qquad\qquad\quad\times ~~
\Big(
{1\over 2\pi }\int\limits_{t'_2=0} dt'_2 \p_t X_{++}(t'_2) \left ( {t'_2}^N(t'_2+a)^M \right ) ^{-{m\over N}}\Big )
|0\rangle_t \;.\qquad\qquad
\label{ptfiveqq2}
\eea
This expression is the same as (\ref{ptfiveq}) with the replacements
\be
M\r N ~~~N\r M~~~ a\r -a \,.
\ee
Thus $f^{B(2)}_{mk}$ can be obtained from (\ref{exactft}) with the above replacements.  For ${k \over M+N} \neq {m\over N}$ this gives:
\be
f^{B(2)}_{mk}~=~
{(-1)^m \sin(\pi {Nk\over M+N})\over \pi (M+N)}\;
{a^{k-{(M+N)m\over N}}\over {k\over M+N}-{m\over N}}\;
{\Gamma[{(M+N) m\over N}]\over \Gamma[m]\Gamma[{Mm\over N}]}\,
{\Gamma[{Nk\over M+N}]\Gamma[{Mk\over M+N}]\over \Gamma[k]}
\label{fpreq}
\ee
and we have the special case
\be
f^{B(2)}_{mk}|_{{k\over M+N}={m\over N}}~=~{N\over M+N} \,.
\ee
We can again use fractional modes
\be
r={m\over N}, ~~~s={k\over M+N}
\ee
 When expressing our result using $r$ and $s$ in place of $m$ and $k$, we will write
 \be
 f^{B(2)}_{mk}\r \t f^{B(2)}_{rs}\,.
 \ee
We then have for $r \neq s$:
 \bea
\t f^{B(2)}_{rs}&=&
-{i\over 2\pi}
z_0^{s-r}
{\mu_s\over s-r}\,
{1\over  (M+N)}\left( {(M+N)^{M+N}\over M^MN^N}\right)^{(s-r)}
{\Gamma[(M+N)r]\over \Gamma[Mr]\Gamma[Nr]}\,
 {\Gamma[Ms]\Gamma[Ns]\over \Gamma[(M+N)s]}\nn
 \label{exactf2}
\eea
while for $r = s$:
\be
\t f^{B(2)}_{rs}|_{r=s} = {N \over M+N}\,.
\ee

\subsection{ Summarizing the result for $\t f^{B(1)}_{qs}$, $\t f^{B(2)}_{rs}$}
 
For convenient reference, we record the result for $\t f^{B(1)}_{qs}$, $\t f^{B(2)}_{rs}$ for general $M,N$:
\bea
\t f^{B(1)}_{qs} & = & \begin{cases}
{M \over M+N} & q = s \\
{i\over 2\pi}z_0^{s-q}
{\mu_s \over s-q}\,
{1\over  (M+N)}\left( {(M+N)^{M+N}\over M^MN^N}\right)^{(s-q)}
{\Gamma[(M+N)q]\over \Gamma[Mq]\Gamma[Nq]}\,
{\Gamma[Ms]\Gamma[Ns]\over \Gamma[(M+N)s]}
\qquad\quad & q \neq s 
\end{cases}\qquad\quad
\eea

\bea
\t f^{B(2)}_{rs} & = & \begin{cases}
{N \over M+N} & r = s \\
-{i\over 2\pi}z_0^{s-r}
{\mu_s\over s-r}\,
{1\over  (M+N)}\left( {(M+N)^{M+N}\over M^MN^N}\right)^{(s-r)}
{\Gamma[(M+N)r]\over \Gamma[Mr]\Gamma[Nr]}\,
 {\Gamma[Ms]\Gamma[Ns]\over \Gamma[(M+N)s]} \qquad\quad & r \neq s
\end{cases}\qquad\quad
\eea

\section{The continuum limit}\label{sec:cont}

While we have obtained the exact expressions for $\gamma^B_{kl}$, $f^{B(1)}_{qs}, f^{B(2)}_{rs}$, the nature of the physics implied by these expressions may not be immediately clear because the expressions look somewhat involved. We will comment on the structure of these expressions in the discussion section below. But first we note that these expressions simplify considerably in the limit where the arguments $q, r, s$ are much larger than unity. We call the resulting approximation the `continuum limit', since large mode numbers correspond to short wavelengths, and at short wavelength the physics is not sensitive to the finite length of the component string. Thus the expressions in this continuum limit describe the results obtained in the limit of  {\it infinite} component strings. Such expressions are useful for the following reason. When the D1D5 system is used to describe a black hole, then the total winding $N_1N_5$ of the component strings is very large, and so one expects the individual component strings to have large winding as well. Having long component strings ($M, N$ much larger than unity) is approximately  equivalent to holding $M,N$ fixed and taking $q,r,s$ large.

\subsection{Continuum limit for the $\gamma^B_{kl}$}\label{sec:gammacont}

Let us start by looking at $\gamma^B_{kl}$. We have the exact expression
\bea
\t\gamma^B_{ss'}&=&{1\over 4\pi^2}\,z_0^{s+s'}\,
{\mu_s\mu_{s'}\over  s+s'}\,{MN\over (M+N)^3} 
\left ({(M+N)^{M+N}\over M^MN^N}\right )^{s+s'} 
{\Gamma[{Ms}]\Gamma[{Ns}]\over \Gamma[(M+N)s]}~{\Gamma[{Ms'}]\Gamma[{Ns'}]\over \Gamma[(M+N)s']}\,.\nn
\label{wnine}
\eea
We wish to find an approximation for this expression when
\be
s\gg 1, ~~~s'\gg 1
\ee
We have the basic identity, for positive integer $K$:
\be
\Gamma[x]\Gamma[x+{1\over K}]\Gamma[x+{2\over K}]\dots \Gamma[x+{K-1\over K}]=(2\pi)^{K-1\over 2}K^{\h-Kx}\Gamma[Kx]
\ee
Using this identity, we get
\be
\Gamma[{Ms}]={1\over (2\pi)^{M-1\over 2}M^{\h-M s}}
\Gamma[s]\Gamma[s+{1\over M}]\dots \Gamma[s+{M-1\over M}]
\ee
\be
\Gamma[{Ns}]={1\over (2\pi)^{N-1\over 2}N^{\h-N s}}
\Gamma[s]\Gamma[s+{1\over N}]\dots \Gamma[s+{N-1\over N}]
\ee
\be
\Gamma[(M+N)s]={1\over (2\pi)^{M+N-1\over 2}(M+N)^{\h-(M+N) s}}
\Gamma[s]\Gamma[s+{1\over M+N}]\dots \Gamma[s+{M+N-1\over M+N}]
\ee
We have, for $s\gg 1$, $x\ll s$
\be
{\Gamma[s+x]\over \Gamma[s]}\approx s^x
\ee
which gives
\be
{\Gamma[s+x]}\approx {\Gamma[s]} s^x
\ee
We use the above approximations in the expression
\be
{\Gamma[{Ms}]\Gamma[{Ns}]\over \Gamma[(M+N)s]}
\ee
There are an equal number of factors ${\Gamma[s]}$ at the numerator and denominator, so they cancel out. We can now collect the powers of $s$. In the numerator we have the power
\be
\left ( {1\over M}+{2\over M}+\dots {M-1\over M} \right ) + \left ({1\over N}+{2\over N}+\dots {N-1\over N}\right ) ={M+N-2\over 2}
\ee
In the denominator we have the power
\be
{1\over N+M}+{2\over N+M}+\dots {N+M-1\over N+M}={(M+N-1)\over 2}
\ee
Thus overall we get
\be
{\Gamma[{Ms}]\Gamma[{Ns}]\over \Gamma[(M+N)s]}\approx {(2\pi)^\h \left ({M^MN^N\over (M+N)^{M+N}}\right )^s \sqrt{M+N\over MN}}\, s^{-\h}
\label{wthree}
\ee
Similarly, we get
\be
{\Gamma[{Ms'}]\Gamma[{Ns'}]\over \Gamma[(M+N)s']}\approx {(2\pi)^\h \left ({M^MN^N\over (M+N)^{M+N}}\right )^{s'} \sqrt{M+N\over MN}}\, {s'}^{-\h}
\ee
Using these approximations in (\ref{wnine}), we find
\be
\t\gamma^B_{ss'}\approx{1\over 2\pi}{1\over(M+N)^2}\,z_0^{s+s'}{\mu_s\mu_{s'}\over \sqrt{ss'}}\,{1\over s+s'}\,.
\ee

\subsection{Continuum limit for the $f^{B(i)}_{qs}$}\label{sec:fcont}

Let us also obtain the continuum limit for the $f^{B(i)}_{qs}$. 
We have the exact expression
 \bea
\t f^{B(1)}_{qs}&=&
{i\over 2\pi}
z_0^{s-q}
{\mu_s\over s-q}\,
{1\over (M+N)}\left( {(M+N)^{M+N}\over M^MN^N}\right)^{(s-q)}
{\Gamma[(M+N)q]\over \Gamma[Mq]\Gamma[Nq]}\,
 {\Gamma[Ms]\Gamma[Ns]\over \Gamma[(M+N)s]} \,. \nn
 \label{exactftq}
 \eea
 We have, from (\ref{wthree})
 \be
{\Gamma[{Ms}]\Gamma[{Ns}]\over \Gamma[(M+N)s]}\approx {(2\pi)^\h \left ({M^MN^N\over (M+N)^{M+N}}\right )^s \sqrt{M+N\over MN}}\, s^{-\h}
\ee
Similarly,
\be
{\Gamma[(M+N)q]\over \Gamma[{Mq}] \Gamma[{Nq}]}\approx {(2\pi)^{-\h} \left ({M^MN^N\over (M+N)^{M+N}}\right )^{-q} \sqrt{MN\over M+N}}\, q^{\h}
\ee
Thus we obtain
\be
\t f^{B(1)}_{qs}\approx {i\over 2\pi}{1\over (M+N)}z_0^{s-q}
\mu_s\sqrt{q\over s}\,{1\over s-q}\,.
\ee
Similarly, for $\t f^{B(2)}_{rs}$ we obtain
\be
\t f^{B(2)}_{rs}\approx -{i\over 2\pi}{1\over (M+N)}z_0^{s-r}
\mu_s\sqrt{r\over s}\,{1\over s-r}\,.
\ee

\section{Discussion}\label{sec:disc}

We have considered the effect of the twist operator $\sigma_2$ when it links together component strings of windings $M,N$ into a single component string of length $M+N$. We have found the final state in the case where the initial state on the component strings was the vacuum $|0\rangle^{(1)}\otimes |0\rangle^{(2)}$, and also in the case where one of the strings had an initial excitation $\alpha^{(i)}_{-m}$.

While we have discussed this problem in the context of the D1D5 CFT, we note that this is a problem that could arise in other areas of physics. Each component string describes a free field theory on a circle, and the twist interaction joins these circles. In this process the vacuum state of the initial field theory goes over to a `squeezed state' of  the final theory; the coefficients $\gamma^B_{ss'}$ describe this squeezed state.

It is interesting to analyze the structure of the results that we have found. Consider the expression for $\gamma^B_{kl}$ given in (\ref{gammapre}). This contains a factor
\be
{1\over \Gamma[k]}{1\over \Gamma[l]}
\ee
which vanishes when either of the integers $k,l$ is zero or negative.  From (\ref{pfive}) we see that this implies that the $\gamma^B_{kl}$ will multiply only creation operators. Similarly, in the expression (\ref{fpre}) for $f^{B(1)}_{mk}$ we have the factor
\be
{1\over \Gamma[m]}{1\over \Gamma[k]}
\label{wel}
\ee
which vanishes when either of the integers $m$ or $k$ is zero or negative. Thus the function $f^{B(i)}_{mk}$ vanishes unless we have creation operators  in the initial state, and it yields  only creation operators in the final state.

A second observation is that the expression for $\gamma^B_{ss'}$  almost separates into a product of terms corresponding to $s$ and $s'$. Only the term ${1\over s+s'}$ fails to separate in this manner. The part corresponding to $s$ contains the beta function ${\Gamma[{Ms}]\Gamma[{Ns}]\over \Gamma[(M+N)s]}$
and the part for $s'$ contains the beta function ${\Gamma[{Ms'}]\Gamma[{Ns'}]\over \Gamma[(M+N)s']}$.

Similarly,  $f^{B(i)}_{qs}$ almost separates into a product of terms corresponding to $q$ and $s$. Only the term ${1\over s-q}$ fails to separate. The part for $s$ has the beta function $ {\Gamma[Ms]\Gamma[Ns]\over \Gamma[(M+N)s]}$, and the part for $q$ has the reciprocal of the beta function ${\Gamma[(M+N)q]\over \Gamma[Mq]\Gamma[Nq]}$. The appearance of the reciprocal here may be attributed to the fact that the index $q$ corresponds to an operator in the initial state while the index $s$ appears in the final state. (In the case of $\gamma^B_{kl}$, both operators appear in the final state.) In  particular, the appearance of the reciprocal of the beta function ensures that we get the $\Gamma$ functions in the form (\ref{wel}) required to ensure that only creation operators are involved in the effect of $f^{B(i)}_{mk}$. 

We have considered only bosonic excitations in this paper. A similar analysis can be done for the fermionic excitations, and the supercharge $G^-$ in (\ref{pert}) should then be applied to the overall state for bosons and fermions. These steps were carried out for the case $M=N=1$ in \cite{acm2,acm3}; for the case of general $M,N$ we hope to present the results elsewhere.

The expressions for $\gamma^B_{ss'}, f^{B(i)}_{qs}$ simplify considerably in the continuum limit. This  limit  may be more useful for obtaining the qualitative dynamics of thermalization, which is expected to be dual to the process of black hole formation. It would therefore be helpful to have a way of obtaining the continuum limit expressions directly, without having to obtain the exact expressions first. We hope to return to this issue elsewhere.

In general, it is hoped that by putting together knowledge of the fuzzball construction 
(which gives the gravity description  of individual black hole microstates) and  dynamical processes in the interacting CFT (which include black hole formation), we will arrive at a deeper understanding of black hole dynamics.

\section*{Acknowledgements}

We thank Steve Avery, Borun Chowdhury, Amanda Peet  and Ida Zadeh for discussions on various aspects of the twist interaction. The work  is
supported in part by DOE grant DE-FG02-91ER-40690.

\end{document}